\documentclass[nonacm,sigplan]{acmart}
\usepackage[capitalise,noabbrev]{cleveref}
\usepackage{subfig}
\newcommand{\ignore}[1]{}
\usepackage{enumitem}
\usepackage{multirow}
\usepackage{multicol}

\usepackage{tikz}
\newcommand*\bcircled[1]{\tikz[baseline=(char.base)]{
            \node[shape=circle,draw,inner sep=1pt,fill=black, text=white] (char) {#1};}}
\newcommand*\circled[1]{\tikz[baseline=(char.base)]{
            \node[shape=circle,draw,inner sep=1pt,fill=white, text=black] (char) {#1};}}
\newcommand*\beigecircled[1]{\tikz[baseline=(char.base)]{
            \node[shape=circle,draw,inner sep=1pt,fill=beige, text=black] (char) {#1};}}

\AtBeginDocument{%
  }

\setcopyright{none} 

\usepackage{lipsum}
\usepackage{amsmath}
\usepackage{xspace}
\usepackage{booktabs}
\usepackage{tabularx}
\usepackage{makecell}
\usepackage{braket}
\usepackage{enumitem}

\usepackage{amsmath}    %
\usepackage{array}      %
\usepackage{multirow}   %
\usepackage{caption}    %

\usepackage{algorithm}
\usepackage{algpseudocode}
\usepackage{threeparttable}

\usepackage{algorithm}%
\usepackage{algpseudocode}%
\usepackage{multirow}
\usepackage{array, xcolor, pifont}

\newcommand{\greentick}{{\color{teal}\large\ding{51}}}
\newcommand{\redcross}{{\color{red}\large\ding{55}}}
\newcommand{\specialcell}[2][c]{%
\begin{tabular}[#1]{@{}c@{}}#2\end{tabular}}

\newcommand{\attackname}{Zero Knowledge Tampering Attack}
\newcommand{\attackacro}{ZKTA}
\newcommand{\design}{QONTEXTS}
\newcommand{\designenhanced}{\design{}+AD}
\newcommand{\designfullname}{\underline{Q}uantum C\underline{ontext} \underline{S}witching}
\newcommand{\algo}{MFCS}

\definecolor{OliveGreen}{HTML}{135D66}
\usepackage{tcolorbox}
\newtcolorbox{hintbox}[2][]
{
  colframe = OliveGreen!100,
  colback  = OliveGreen!5,
  boxsep=1pt,
  boxrule=0.1mm,
  titlerule=0mm,
  width=\dimexpr\columnwidth\relax, 
  coltitle = blue!20!black,
  title    = #2,
  #1,
}
\usepackage{physics}

\usepackage{balance}

\usepackage{enumitem}

\definecolor{beige}{HTML}{F7EEDD}
\usepackage{enumitem}

\usepackage{tikz}

\begin{document}

\title{Context Switching for Secure Multi-programming of Near-Term Quantum Computers}
    
\author{Avinash Kumar}
\email{avinkumar@utexas.edu}
\affiliation{%
  \institution{The University of Texas at Austin}
  \streetaddress{}
  \city{Austin}
  \state{TX}
  \country{USA}
  \postcode{}
}
\author{Meng Wang}
\email{mengwang@ece.ubc.ca}
\affiliation{%
  \institution{The University of British Columbia}
  \streetaddress{}
  \city{Vancouver}
  \state{BC}
  \country{Canada}
  \postcode{}
}
\author{Chenxu Liu}
\email{chenxu.liu@pnnl.gov}
\affiliation{%
  \institution{Pacific Northwest National Lab}
  \streetaddress{}
  \city{Richland}
  \state{WA}
  \country{USA}
  \postcode{}
}

\author{Ang Li}
\email{ang.li@pnnl.gov}
\affiliation{%
  \institution{Pacific Northwest National Lab}
  \streetaddress{}
  \city{Richland}
  \state{WA}
  \country{USA}
  \postcode{}
}

\author{Prashant J. Nair}
\email{prashantnair@ece.ubc.ca}
\affiliation{%
  \institution{The University of British Columbia}
  \streetaddress{}
  \city{Vancouver}
  \state{BC}
  \country{Canada}
  \postcode{}
}

\author{Poulami Das}
\email{poulami.das@utexas.edu}
\affiliation{%
  \institution{The University of Texas at Austin}
  \streetaddress{}
  \city{Austin}
  \state{TX}
  \country{USA}
  \postcode{}
}
\renewcommand{\shortauthors}{Kumar et al.}

\begin{abstract}

Multi-programming quantum computers improve device utilization and throughput. However, crosstalk from concurrent two-qubit CNOT gates poses security risks, compromising the fidelity and output of co-running victim programs. We design \textit{\attackname{}s (\attackacro{}s)}, using which attackers can exploit crosstalk \emph{without} knowledge of the hardware error profile. \attackacro{}s can alter victim program outputs in 40\% of cases on commercial systems.

We identify that \attackacro{}s succeed because the attacker's program consistently runs with the same victim program in a fixed context. To mitigate this, we propose \textit{\design{}}: a context-switching technique that defends against \attackacro{}s by running programs across multiple contexts, each handling only a subset of trials. \design{} uses \emph{multi-programming with frequent context switching} while identifying a unique set of programs for each context. This helps limit only a fraction of execution to \attackacro{}s. We enhance \design{} with \textit{attack detection} capabilities that compare the distributions from different contexts against each other to identify noisy contexts executed with \attackacro{}s. Our evaluations on real IBMQ systems show that \design{} increases program resilience by three orders of magnitude and fidelity by 1.33$\times$ on average. Moreover, \design{} improves throughput by 2$\times$, advancing security in multi-programmed environments.

\end{abstract}

\maketitle

\section{Introduction}

Near-term quantum systems promise computational speedups for many critical application domains, such as optimization, simulations, healthcare, etc.~\cite{ibmqproteinfolding,farhi2014quantum,mcclean2016theory,biamonte2017quantum,peruzzo2014variational}. This has led to increasing demands for quantum resources from both research groups and enterprises. For example, IBM alone provides quantum access to over 210 organizations, including corporations, universities, research labs, and startups~\cite{ibmq_network}. However, the growth in the number of quantum systems has not kept pace, creating a massive \textit{demand versus supply gap} between users and available quantum resources.

Multi-programming bridges this gap by executing multiple programs concurrently on a quantum system~\cite{das2019case,liu2021qucloud,niu2023enabling,niu2022parallel,liu2021qucloud,ohkura2022simultaneous,niu2022multi,10.1145/3470496.3527434,mineh2023accelerating,resch2021accelerating,park2023quantum}. As quantum programs can only execute a limited number of gates before encountering errors~\cite{sycamoredatasheet,IBMQ}, it is often impractical for programs to use all available qubits on near-term quantum machines. Multi-programming efficiently uses idle qubits, increasing throughput and reducing wait times. Today, a limited form of multi-programming is already supported on commercial systems from QuEra~\cite{AWS_Quera_parallel_circuits}. However, multi-programming faces security challenges due to crosstalk, where undesired quantum interactions occur between co-running programs~\cite{craik2017high, guerreschi2018two, lienhard2019microwave, neill2018blueprint, reagor2018demonstration, murali2020software, murali2019noise, sarovar2020detecting}. Prior research indicates that attackers can exploit this to lower the fidelity of co-running programs~\cite{ash2020analysis, saki2021qubit, deshpande2023design, phalak2021quantum}. Our paper observes similar results on commercial IBMQ systems and aims to provide low-cost secure multi-programming solutions. 

\begin{figure*}[htp]
    \centering
\includegraphics[width=\textwidth]{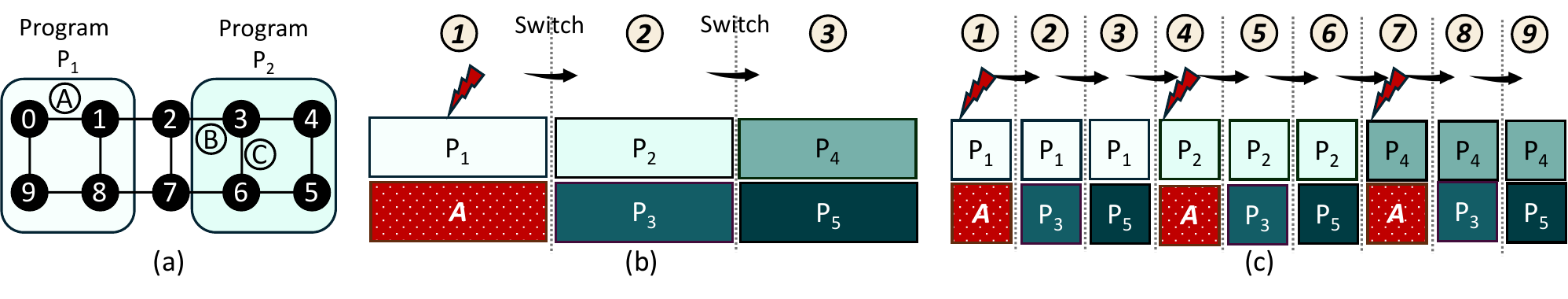} \vspace{-0.25in}
    \caption{(a) A multi-programmed quantum system. (b) Existing solutions always runs the same set of programs together (context) for all trials. (c) \design{} runs each program over many contexts, each with a unique program for a subset of trials. } 
\label{fig:intro}
    \vspace{-0.05in}
\end{figure*}

    \begin{table*}[!htb]
        \centering
        \begin{center}
        \caption{Comparison of multi-programming approaches and their security against \textit{non-local crosstalk-based} attacks}
        \vspace{-0.1in}
        \label{tab:summaryofpriorworks}
        \renewcommand{\arraystretch}{1.0}
         \resizebox{\textwidth}{!}{
        \begin{tabular}{|c|c|c|c|c|c|c|}
            \hline
            Technique & Approach & \specialcell{Does Not\\Need Profiling} & Scalable & \specialcell{Maximum\\Throughput} & \specialcell{Maximizes\\Utilization} & \specialcell{Secure Against\\\emph{Non-Local} Crosstalk Attacks}\\
            \hline
            Multi-programming (MP)~\cite{das2019case} & \specialcell{Fair resource allocation} & \greentick & \greentick & \greentick \phantom{ } ($2\times$)& \greentick & \redcross \\
            \hline\specialcell{ QuCloud+~\cite{liu2024qucloud+}} & \specialcell{Crosstalk-aware scheduling} & \redcross & \greentick &  \greentick \phantom{ } ($2\times$) & \greentick & \redcross\\
            \hline
            QuMC~\cite{niu2023enabling} & \specialcell{Isolate programs}& \redcross & \greentick & \greentick \phantom{ } ($2\times$) & \redcross & \redcross\\
            \hline
            \specialcell{Antivirus~\cite{deshpande2023design}} & \specialcell{Detect and deny execution}& \greentick &  \redcross & \redcross \phantom{ } ($<2\times$) & \redcross & \redcross \\
            \hline
            \textbf{\design{} (Ours)} & \specialcell{Context switching}& \greentick & \greentick & \greentick \phantom{ } ($2\times$) & \greentick & \greentick \\
            \hline
        \end{tabular}
        }
        \end{center}
        \vspace{-0.05in}
    \end{table*}

Practical quantum programs rely on hundreds to thousands of CNOT gates for generating entanglement, a critical mechanism for achieving quantum speedup~\cite{basso2021quantum,bian2019quantum,nam2020ground,lolur2021benchmarking,kim2021universal,guerreschi2019qaoa}. Crosstalk between concurrent CNOT operations poses a significant interference challenge in multi-programmed systems. For instance, in Figure~\ref{fig:intro}(a), while running two programs, $P_1$ and $P_2$, any crosstalk from CNOTs in $P_2$ can increase error rates of CNOTs crucial to $P_1$, thereby reducing its fidelity. High-crosstalk pairs, such as CNOTs \circled{A} and \circled{B}, significantly elevate the error rates. Our experiments on 27-qubit IBM Hanoi reveal that even non-neighboring links (such as CNOTs \circled{A} and \circled{C}) can diminish program fidelity by up to 18\%. We demonstrate that an attacker can exploit such high-crosstalk links to completely manipulate the output of a victim program, even without prior knowledge of the machine's crosstalk profile. This exploit strategically executes a large number of CNOTs to maximize the activation of high-crosstalk links and intensify crosstalk effects.

Prior works attempt to defend against such attacks. But they either lack security, reduce utilization, or need extensive profiling, as summarized in Table~\ref{tab:summaryofpriorworks}. For example, QuCloud+ profiles the machine to identify high-crosstalk link pairs and avoid parallel CNOTs on them~\cite{liu2024qucloud+}. As profiling crosstalk between all possible link pairs scales exponentially, compilers \emph{only} profile local pairs one hop away from each other~\cite{murali2020software}. So, they overlook non-local high-crosstalk links which reduces security. In  Figure~\ref{fig:intro}(a), QuCloud+ avoids scheduling CNOTs \circled{A}, and \circled{B} concurrently but will schedule \circled{A} and \circled{C} in parallel, even if they form a high-crosstalk pair.

QuMC~\cite{niu2023enabling, niu2022parallel} attempts to mitigate crosstalk by isolating programs, leaving a layer of unused qubits between them, such as qubits 2 and 7 in Figure~\ref{fig:intro}(a). However, like QuCloud+, QuMC cannot handle non-local high-crosstalk link pairs. Also, leaving qubits unused reduces utilization. Another approach detects attack circuits via pattern matching~\cite{deshpande2023design} and denies their execution. This method relies on an NP-complete graph isomorphism algorithm that does not scale to large programs, limiting its practicality. %
\textit{Our goal is to enable scalable and secure multi-programming, without reducing throughput.}

\vspace{0.05in}
\noindent \textbf{Insights on Attack Generation:} We propose the \textit{\attackname{} (\attackacro{})} to better understand these insecurities. \attackacro{} uses two key insights. First, even without knowing the system's crosstalk profile, the attacker \emph{predicts} that any CNOT in their program may form a non-local high-crosstalk pair with an ongoing CNOT in the victim program. Second, it uses our studies on commercial systems that show crosstalk increases with the number of parallel CNOTs. The \attackacro{} executes as many concurrent CNOTs as possible in each cycle to maximize the probability of a successful attack. By activating a large number of links every cycle, the \attackacro{} (1) increases the probability of forming a high-crosstalk pair with ongoing CNOTs in the victim program and (2) amplifies crosstalk. We demonstrate \attackacro{}s on real IBMQ machines and, in a case study, show that \attackacro{}s tamper with the victim program's output in 40\% of cases. More importantly, we show that \attackacro{}s \emph{can be camouflaged} as benign programs by leveraging specific CNOT patterns.

\vspace{0.05in}
\noindent \textbf{Low-Cost Defense:} We define a \emph{context} as a set of co-located programs executed together in a multi-programmed system. The vulnerability to \attackacro{}s arises from leveraging consistent contexts throughout the duration of program execution. For example, Figure~\ref{fig:intro}(b) illustrates three contexts in a multi-programmed system, where program $P_1$ runs with an attacker's program $A$ for all trials in the first context. Only after [$P_1, A$] completes execution does the context switch to run program pair [$P_2, P_3$]. This approach allows an attacker to consistently degrade the fidelity of the co-located victim program such as $P_1$, while $P_2$ to $P_5$ remain unaffected. We propose \textit{\design{}: \underline{Q}uantum C\underline{ontext} \underline{S}witching} to mitigate this. \design{} leverages the insight that running each program across multiple contexts can defend against \attackacro{}s because in this approach, each context executes only a subset of the trials with unique programs. This exposes only a fraction of the execution to potential attacks. 

To this end, \design{} uses \textit{Multi-programming with Frequent Context Switching (\algo{})} algorithm. \algo{} assigns a unique set of programs to each context \emph{without} requiring any profiling. Figure~\ref{fig:intro}(c) illustrates \design{}, where each program runs across three contexts. Each context executes one-third of the trials with a unique program selected by \algo{}. For instance, $P_1$ runs in contexts \beigecircled{\textbf{\textit{1}}}, \beigecircled{\textbf{\textit{2}}}, and \beigecircled{\textbf{\textit{3}}} with $A$, $P_3$, and $P_5$, respectively. This limits $P_1$'s exposure to $A$ to one-third, leaving the remaining two-thirds unaffected. \design{} only dynamically reduces the length of contexts and does not alter the total number of trials per program.

We further enhance \design{} with the \textit{Hold-Out method} to detect \attackacro{}s. We call this method \design{} with Attack Detection or {\em \designenhanced{}}. It estimates noise levels in each context by comparing the distributions from different contexts via statistical measures, like Hellinger distance~\cite{hellingerdistance}. Contexts with \attackacro{}s exhibit noisy, inaccurate distributions that significantly differ from those run with benign programs. The trials from  attacked contexts are discarded, and the results from the remaining trials are aggregated.

\vspace{0.05in}
\noindent \textbf{Contributions:} This paper makes four key contributions.

\begin{enumerate}[leftmargin=0cm,itemindent=.5cm,labelwidth=\itemindent,labelsep=0cm,align=left, itemsep=0.05 cm, listparindent=0.5cm]
    \item Demonstrates that crosstalk between non-local CNOT gates significantly increases their error rates.
    \item Introduces \textit{\attackname{} (\attackacro{})}, that exploits crosstalk between \emph{non-neighboring} CNOTs to degrade the fidelity and tamper with the output of co-running programs in multi-programmed machines. 
    \item Proposes \textit{\design{}: \underline{Q}uantum C\underline{ontext} \underline{S}witching}, which defends against \attackacro{}s by co-locating each program with various other programs across multiple contexts.
    \item Develops \textit{\designenhanced{}}, which compares context distributions to identify and eliminate \attackacro{}s. 
\end{enumerate}

Our studies on state-of-the-art IBMQ systems show that \design{}: (1) defends against \attackacro{}s, (2) improves resilience by three orders of magnitude, (3) achieves 2$\times$ throughput as default multi-programming, (4) improves fidelity by 1.33$\times$ on average compared to multi-programming, and (5) attains fidelity of isolated mode in the best-case. \design{} remains effective even for large programs, multiple systems, increased concurrency, and across calibration cycles. 

\section{Background}
\subsection{Multi-programming in Quantum Computers}

Noisy devices limit programs from using all available qubits on near-term systems. Table~\ref{tab:qv} shows the utilization of some recent systems based on their quantum volume, a measure of the largest square circuit of random two-qubit gates a system can successfully run~\cite{quantumvolume}. A QV of 512 on IBM Prague means it can reliably run circuits with up to 9 qubits and 9 layers of random CNOTs, thereby using only 33.3\% of the qubits. Simultaneously, the number of quantum users globally far exceeds the number of available systems, creating a huge gap between them, often noticeable as long wait times ranging from a few hours to  days~\cite{ionq_cloud,IBMQ,rigetti_qcs}. Multi-programming bridges the gap by efficiently using idle qubits to run multiple programs concurrently. Thus, the system throughput and utilization increase, leading to shorter wait times for users.

    \begin{table}[!htb]
           \vspace{-0.1in}
        \centering
        \begin{center}
        \caption{Utilization for Quantum Volume (QV) Circuits}
        \vspace{-0.1in}
        \label{tab:qv}
        \renewcommand{\arraystretch}{1.0}
        \resizebox{\columnwidth}{!}{
        \begin{tabular}{|c|c|c|c|}
            \hline
            Machine & \#Qubits & QV & Utilization (\%)\\
            \hline
            \hline
            IBM Montreal & 27 & 128~\cite{montrealqv} & 25.9\\
            \hline
            IBM Prague & 27 & 512~\cite{pragueqv} & 33.3\\
            \hline
            AQT Pine & 24 & 128~\cite{aqtpineqv} & 29.2\\
            \hline
            Quantinuum H2 & 56 & 262K~\cite{quantinuumh2qv} & 32.1 \\
            \hline
        \end{tabular}
        }
        \end{center}
    \end{table}

\ignore{
\subsection{Multi-programming bridges users to systems gap}
The promise of computational advantage and availability of hardware has led to increased demands for quantum computing resources in recent years from both research groups that want to advance the technology and enterprises that are evaluating the potential of quantum technologies for addressing their business needs. However, the number of quantum systems available has not grown at the same pace, leading to a significant demand-to-supply gap between users and device providers. For example, IBM provides quantum resource access to 210+ organizations, including several Fortune 500 companies, universities, laboratories, and startups~\cite{ibmq_network}. 

Multi-programming bridges this gap by running multiple programs concurrently. Multi-programming improves the throughput of quantum computers by reducing the number of idle qubits at any given time~\cite{das2019case}. For example, in Figure~\ref{fig:backgrounda}, running programs $P_1$ and $P_2$ concurrently on the 27-qubit IBM Hanoi increases the throughput and utilization by 2$\times$. 
}

Quantum devices exhibit variable error rates~\cite{murali2019noise,tannu2018case,dasgupta2021stability}. Thus, in shared systems, programs may be forced to use inferior devices. Prior works address this through intelligent resource partitioning and instruction scheduling, ensuring programs in shared settings are allocated similar quality devices to those they would use in isolation~\cite{das2019case,niu2022multi,niu2023enabling,niu2022parallel,liu2021qucloud,ohkura2022simultaneous,liu2024qucloud+}.

\subsection{Security Concerns in Multi-programmed Systems}
The fidelity of programs decreases due to crosstalk, which occurs when undesired quantum interactions are activated during operations. Prior works show that crosstalk from concurrent CNOTs reduce the fidelity of multi-programmed systems~\cite{ash2020analysis,deshpande2023design}. This is exploited to develop crosstalk-based attacks. Our studies on IBMQ systems (Section~\ref{sec:bvattacked} and~\ref{sec:zktaonvqas}) show that such attacks can alter program outputs \emph{without} prior knowledge of hardware errors. Even worse, these attacks can be camouflaged as benign programs (Section~\ref{sec:benignapps}).

\subsection{Crosstalk Vulnerabilities: A Grim Reality}
Crosstalk-based vulnerabilities exploit fundamental device-level imperfections. For example, superconducting qubits are controlled by microwave tones sent through cavities, which can affect unintended qubits. Residual coupling between qubits cause such unwanted interactions, leading to crosstalk. Specifically, interactions between superconducting qubits via microwave resonators cause ZZ couplings. While tunable couplers suppress these interactions to some extent~\cite{PhysRevApplied.12.054023}, residual qubit-qubit coupling and frequency collisions still result in crosstalk~\cite{PRXQuantum.3.020301}. As the range of valid frequencies is limited, crosstalk from frequency collisions are unavoidable as systems scale. Crosstalk also exists in other systems, such as trapped ions and neutral atoms, due to unwanted qubit-qubit and qubit-control coupling~\cite{PhysRevLett.129.240504,wang2016single,PhysRevLett.114.100503,PhysRevLett.123.230501}. Consequently, crosstalk is a key source of errors that cannot be fully eliminated at the device-level across real quantum systems and remains a potential security vulnerability.

\subsection{Limitations of Prior Defenses}
Prior works that attempt to defend against such crosstalk-based attacks face one or more of the following issues: 

\subsubsection{Inadequate security} 
\begin{figure}[htp]
    \centering
\includegraphics[width=0.9\columnwidth]{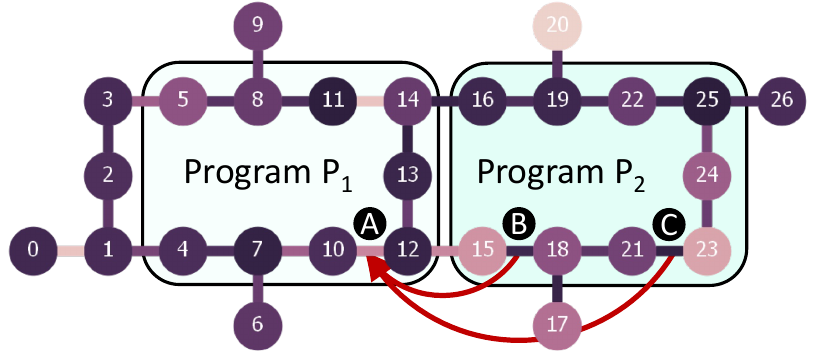} \vspace{-0.1in}
    \caption{Multi-programming improves system utilization, but programs are vulnerable to crosstalk-based attacks. }  \vspace{-0.15in}
\label{fig:backgrounda}
\end{figure}

QuCloud+ employs a limited form of crosstalk-aware scheduling by profiling pairs of links one hop from each other and avoiding concurrent CNOTs on them~\cite{liu2024qucloud+}. For instance, in Figure~\ref{fig:backgrounda}, $Q_{10}\leftrightarrow Q_{12}$ and $Q_{15}\leftrightarrow Q_{18}$ form a one-hop pair. However, QuCloud+ cannot tolerate crosstalk between non-local links, which our studies show to be substantial (Section~\ref{sec:attackdemo}). Even if CNOTs \bcircled{A} and \bcircled{B} are not scheduled in parallel, an attacker ($P_2$), can still exploit non-local high-crosstalk pairs, such as CNOTs \bcircled{A} and \bcircled{C}, against victim ($P_1$). Additionally, the profiling overheads in QuCloud+ scales quadratic in the number of links because the number of combinations or link pairs is $\binom{L}{2}$ for $L$ links. 

\subsubsection{Poor scalability} Antivirus detects attack circuits via pattern matching~\cite{deshpande2023design} and refuses to run them.  However, it relies on an NP-complete graph algorithm that does not scale to large programs, limiting its practical adoption. Also, it cannot handle attack circuits disguised as benign programs and leads to denial of services for such cases. 

\subsubsection{Low utilization} QuMC~\cite{niu2023enabling, niu2022parallel} isolates programs by sparing a layer of qubits between them. For example, $P_2$ avoids qubits $Q_{15}$ and $Q_{16}$, and utilizes $Q_{17}$ and $Q_{20}$ instead, thereby isolating it from $P_1$. However, similar to QuCloud+, QuMC too cannot handle non-local high-crosstalk link pairs, and lowers system utilization. %

\subsection{Threat Model}
Our threat model assumes an attacker in a multi-programmed system aims to degrade the fidelity and possibly tamper with the output of co-running programs. The attacker does not need to know the system's crosstalk characteristics and performs no profiling. We assume the system isolates co-running applications and avoids scheduling parallel CNOTs on links one hop away from each other. We assume \emph{a more practical threat-model} than QuCloud+~\cite{liu2024qucloud+}, QuMC~\cite{niu2023enabling, niu2022parallel}, and Antivirus~\cite{deshpande2023design}. This is because, unlike prior works, we aim to use links that are two or more hops away (non-local links) to induce crosstalk-based attacks.

\section{\attackname{}s}
\label{sec:attackdemo}
We propose the \textit{\attackname{} (\attackacro{})} that degrades fidelity and even tampers with the output of co-running programs in multi-programmed quantum systems. To evaluate the feasibility of \attackacro{}s and explain the intuition behind their formulation, we discuss some crosstalk characterization studies on state-of-the-art IBMQ systems. These systems employ tunable coupling and sparse heavy-hexagonal topologies to maximally reduce crosstalk at device-level. 
Note that we include these studies only to highlight the severity of crosstalk-based attacks and provide intuition. In practice, \attackacro{}s do not need crosstalk profiles to succeed.

\subsection{Insights of \attackacro{}s}
\begin{figure}[htp]
    \centering
\includegraphics[width=\columnwidth]{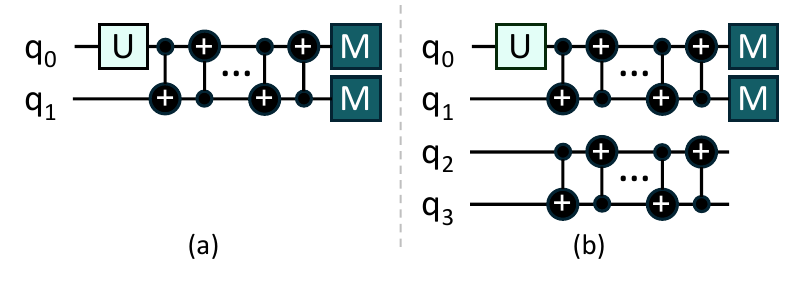} \vspace{-0.2in}
    \caption{Micro-benchmarks (a) $\mu_{b1}$ and (b) $\mu_{b2}$ to profile crosstalk on IBM systems. CNOTs between $q_2$ and $q_3$ in $\mu_{b2}$ generate crosstalk and fidelity of $\mu_{b1}$ is compared with $\mu_{b2}$.}
\label{fig:microbenchmark}
\vspace{-0.1in}
\end{figure}

To evaluate the severity of crosstalk, we use micro-benchmarks shown in Figure~\ref{fig:microbenchmark}. The first one, $\mu_{b1}$, prepares qubit $q_{0}$ in an arbitrary state and performs CNOTs between qubits $q_{0}$ and $q_{1}$. The second one, $\mu_{b2}$, mirrors $\mu_{b1}$ but includes extra CNOTs between qubits $q_{2}$ and $q_{3}$ to induce crosstalk, thereby increasing the error rate of the CNOTs between $q_{0}$ and $q_{1}$. We run these micro-benchmarks on all 682 possible link pairs of 27-qubit IBMQ Hanoi.

\subsubsection{\textbf{Observation-1: Non-Local Crosstalk is Prominent}} 
\begin{figure}[htp]
    \centering\vspace{-0.05in}
\includegraphics[width=\columnwidth]{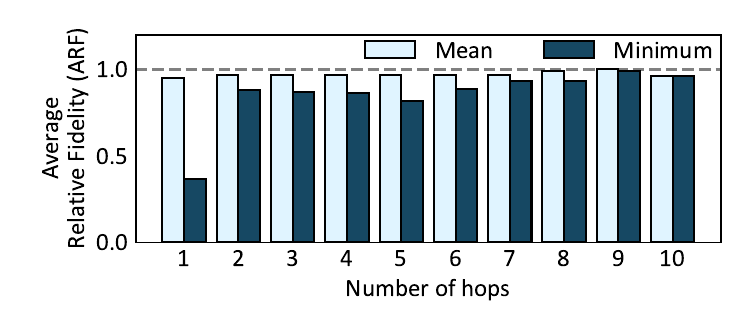} \vspace{-0.25in}
    \caption{Increased number of links used for CNOTs heightens crosstalk and degrades the RF of the micro-benchmarks.}
\label{fig:longrangeseverity}
\vspace{-0.15in}
\end{figure}

We measure the impact of crosstalk using the \textit{Relative Fidelity (RF)} of $\mu_{b1}$, defined as the ratio of fidelity of $\mu_{b2}$ to that of $\mu_{b1}$. We compute Fidelity by comparing the output distribution from real hardware against an error-free one. An RF of 1 means no impact, while an RF below 1 implies increased error rates due to crosstalk. We observe that RF of $\mu_{b2}$ is below 1 for 58.7\% of the link pairs. Figure~\ref{fig:longrangeseverity} shows the mean RF versus hop distance $d$, which denotes the minimum distance between two links in a pair. We observe considerable crosstalk between neighboring link pairs ($d=1$), similar to prior works~\cite{murali2020software}.  However, we also observe substantial crosstalk for non-local links pairs that are distant from each other. For example, CNOTs between two links that are 5-hops away reduce fidelity by up-to 18\%.

\begin{figure}[htp]
\vspace{-0.05in}
    \centering
\includegraphics[width=\columnwidth]{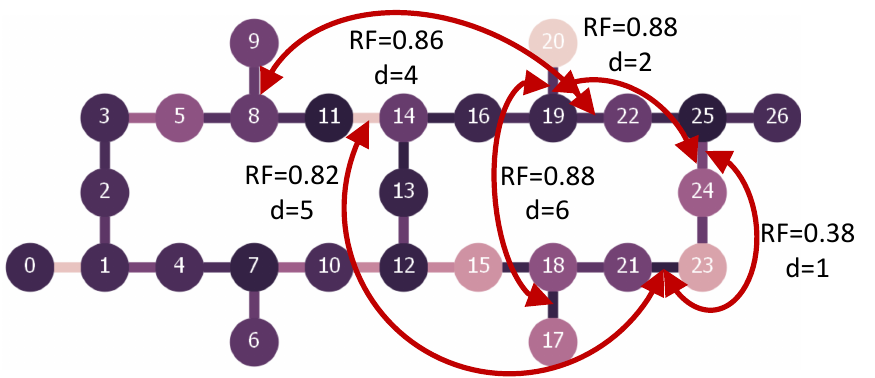} \vspace{-0.2in}
    \caption{High crosstalk pairs on IBM Hanoi shows substantial crosstalk exists even between non-neighboring links.} %
\label{fig:xtalkmap}
\end{figure}

\noindent Figure~\ref{fig:xtalkmap} shows some high crosstalk pairs on IBM Hanoi, highlighting the prominence of the non-local crosstalk even on state-of-the-art IBMQ systems. We observe similar trends even on other IBMQ machines (IBM Sherbrooke, IBM Kyiv, IBM Osaka, IBM Brisbane).

\noindent \textbf{\textit{Insight-1:}} \textit{An \textit{attacker} can exploit non-local CNOTs on distant links far from a co-running \textit{victim} program \textit{guessing} it will likely form a high-crosstalk pair with their ongoing CNOTs.}

\subsubsection{\textbf{Observation-2: Scaling CNOTs Increase Crosstalk}}
\begin{figure}[htp]
    \centering
\includegraphics[width=\columnwidth]{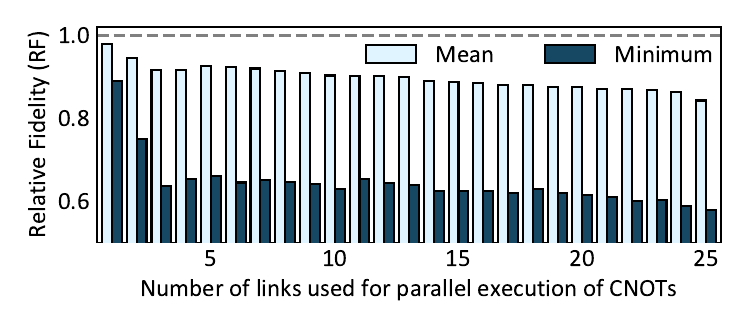} \vspace{-0.25in}
    \caption{Increased number of links used for CNOTs heightens crosstalk and degrades the RF of the micro-benchmarks.}
\label{fig:xtalkamplify}
\vspace{-0.05in}
\end{figure}

To study the impact of increased CNOT concurrency, we modify the second microbenchmark and increase the number of links used to run parallel CNOTs. 
For example, if a $\mu_{b1}$ for IBM Hanoi (Figure~\ref{fig:backgrounda}) uses link $Q_0\leftrightarrow Q_1$, we create two variants of $\mu_{b2}$: one with parallel CNOTs on $Q_0\leftrightarrow Q_1$ and $Q_2\leftrightarrow Q_3$, and another with CNOTs on $Q_0\leftrightarrow Q_1$, $Q_2\leftrightarrow Q_3$, and $Q_4\leftrightarrow Q_7$. We prepare more variants by further increasing the number of links to amplify crosstalk. Figure~\ref{fig:xtalkamplify} shows the mean and minimum RF based on number of links used.

The RF decreases substantially with an increasing number of links activated. For example, executing CNOTs on the link pair $Q_{18}\leftrightarrow Q_{21}$ and $Q_{23}\leftrightarrow Q_{24}$ reduces the RF to 0.89. Adding a third link, $Q_{12}\leftrightarrow Q_{15}$, $Q_{13}\leftrightarrow Q_{14}$, and $Q_{18}\leftrightarrow Q_{17}$, reduces the RF to 0.75. Executing CNOTs on a quadruple of links, $Q_{23}\leftrightarrow Q_{24}$, $Q_{22}\leftrightarrow Q_{25}$, $Q_{25}\leftrightarrow Q_{26}$, and $Q_{18}\leftrightarrow Q_{21}$, further lowers RF to 0.64.

\vspace{0.05in}
\noindent \textbf{\textit{Insight-2:}} \textit{To improve probability of success, the attacker runs as many concurrent CNOTs as possible to maximize the chance of executing CNOTs forming a high crosstalk pair (or triplet and beyond) and crosstalk amplification in the victim program.}

\subsection{Demonstrating \attackacro{}s on IBMQ Systems}
\label{sec:bvattacked}

\begin{figure}[htp]
    \centering %
\includegraphics[width=\columnwidth]{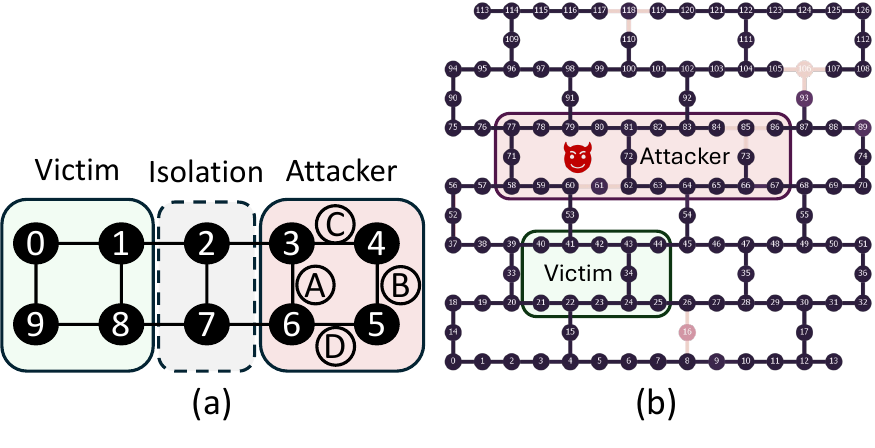} \vspace{-0.3in}
    \caption{(a) Example of a \attackname{} circuit (b) Qubit regions allocated to the victim and attacker while multi-programming on 127-qubit IBM Osaka.} \vspace{-0.1in}
\label{fig:attack_osaka}
\end{figure}

Figure~\ref{fig:attack_osaka}(a) shows a \attackacro{}. The attacker is isolated from the victim program, as in prior works QuMC and QuCloud+~\cite{ash2020analysis,liu2024qucloud+}, and can only employ non-local crosstalk. 
The attacker continuously alternates between cycles of CNOTs- CNOTs \circled{A} and \circled{B} in one cycle, followed by CNOTs \circled{C} and \circled{D} in the next. Figure~\ref{fig:attack_osaka}(b) shows the attacker and victim on IBM Osaka. The victim is a 9-qubit Bernstein Vazirani (BV) program~\cite{bernstein1993quantum} encoding secret \textit{11101011}. The attack is considered successful if the secret can be correctly identified when the BV program executed in isolation but it cannot be determined during multi-programming. As BV programs have one only correct answer, we infer the peak of the distribution as the output~\cite{murali2019noise,tannu2018case}. Our studies on IBM Osaka show that this \textbf{\textit{\attackacro{} can be executed successfully}}, and the output of the BV program while multi-programming is \textit{11101\color{red}{\underline{\textbf{1}}}\color{black}11}, which is \textit{incorrect}. In isolation, the correct string appears with a 14\% probability, whereas in multi-programming, the incorrect string is the dominant output appearing with a 10\% probability. 

\begin{figure}[htp]
    \centering
   \vspace{-0.15in}
\includegraphics[width=\columnwidth]{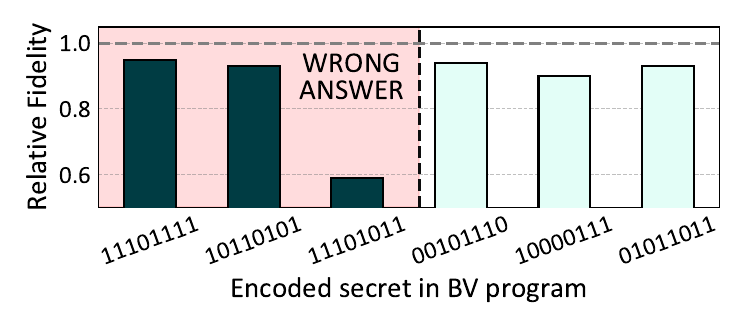} \vspace{-0.25in}
    \caption{Relative Fidelity of Bernstein Vazirani (BV) programs executed concurrently with \attackacro{}s.} 
\label{fig:bvattacks} \vspace{-0.1in}
\end{figure}

Repeating the process with ten unique secret strings show that 40\% of the attacks succeed. Figure~\ref{fig:bvattacks} shows the relative fidelity of some of these BV programs. The three cases on the right refer to scenarios where the BV output can still be inferred despite reduced fidelity, whereas the three cases on the left denote scenarios where the \attackacro{} completely tampers with the correct output.  %

\ignore{
\begin{hintbox}{\color{white}\textbf{Multi-programmed systems are vulnerable}}
An adversary can leverage non-local crosstalk (without actual knowledge of the device error characteristics) to degrade the co-running program(s) fidelity in multi-programmed quantum environments. 
\end{hintbox}}

\noindent \textbf{Generalization of \attackacro{}s:} We study \attackacro{}s by (1)~increasing program sizes, (2)~number of concurrent programs, (3)~using multiple quantum systems, and (4)~across calibration cycles. We observe that \attackacro{}s remain successful in all these scenarios. Our studies also show that \attackacro{}s degrade the performance of promising near-term quantum algorithms, called variational quantum algorithms~\cite{farhi2014quantum,mcclean2016theory}, that use the expectation value of the output distributions (Section~\ref{sec:zktaonvqas}).

\subsection{Disguising \attackacro{}s as Benign Applications}
\label{sec:benignapps}
\begin{figure}[htp]
    \centering
    \vspace{-0.1in}
\includegraphics[width=\columnwidth]{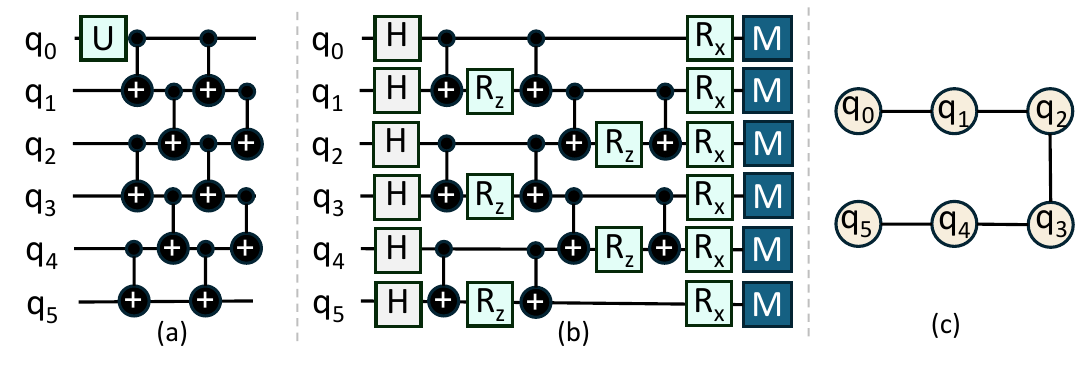} \vspace{-0.2in}
    \caption{(a) The CNOT structure of a \attackacro{} program is translated into (b) a QAOA program by retaining the CNOT structure and introducing the required single-qubit operations. (c) The corresponding graph for MaxCut problem. } \vspace{-0.1in}
\label{fig:zkatoqaoa}
\end{figure}

We explore the feasibility that a \attackacro{} can be disguised as a benign program by exploiting structural similarities. For this, we map the \attackacro{} programs to \textit{Quantum Approximate Optimization Algorithm (QAOA)~\cite{farhi2014quantum}} programs for MaxCut problems. Given a problem graph, QAOA maps each node to a qubit and each edge to RZZ operations on the respective qubits. An RZZ operation involves two CNOTs and a single-qubit $R_Z$ gate. To convert a \attackacro{} program into a QAOA program, we translate its CNOT structure into the RZZ structure of QAOA. Next, we add the required $R_Z$ gates and check if the generated QAOA program maps to a valid graph. If a problem graph can be successfully constructed, an attacker can hide behind such programs, and the device provider cannot detect or refuse to execute them~\cite{deshpande2023design}. Figure~\ref{fig:zkatoqaoa} shows an overview of the process.

\begin{figure}[htp]
    \centering \vspace{-0.1in}
\includegraphics[width=\columnwidth]{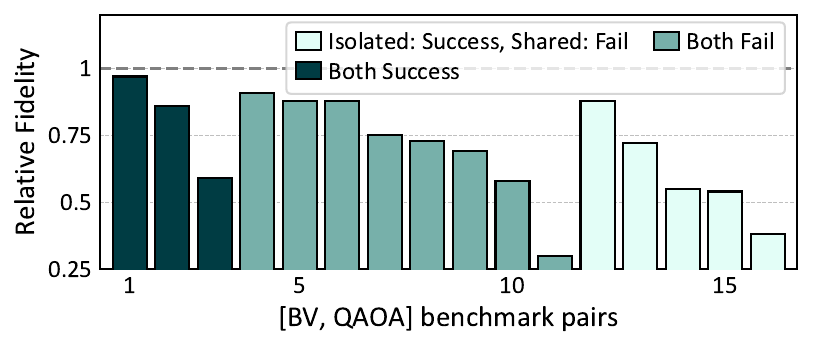} \vspace{-0.2in}
    \caption{Relative Fidelity of BV programs when executed with QAOA programs carefully crafted  from \attackacro{}s.} \vspace{-0.2in}
\label{fig:bvwithqaoaattack}
\end{figure}

We run 16 pairs of such [BV, QAOA] programs on IBM Hanoi. Figure~\ref{fig:bvwithqaoaattack} shows the fidelity of the BV programs relative to their isolated executions. In 3 out of 16 cases (18.75\%), the correct answer can be inferred in both baseline and shared modes. In 8 out of 16 cases (50\%), the correct answer cannot be inferred in both baseline and shared modes. For 5 out of 16 cases (31.25\%), the answer can be inferred correctly in the baseline but not in shared mode, which mean an attacker successfully alters the BV program output by carefully crafting \attackacro{}s that mimic benign QAOA programs. %

\begin{figure*}[htp]
    \centering
\includegraphics[width=\textwidth]{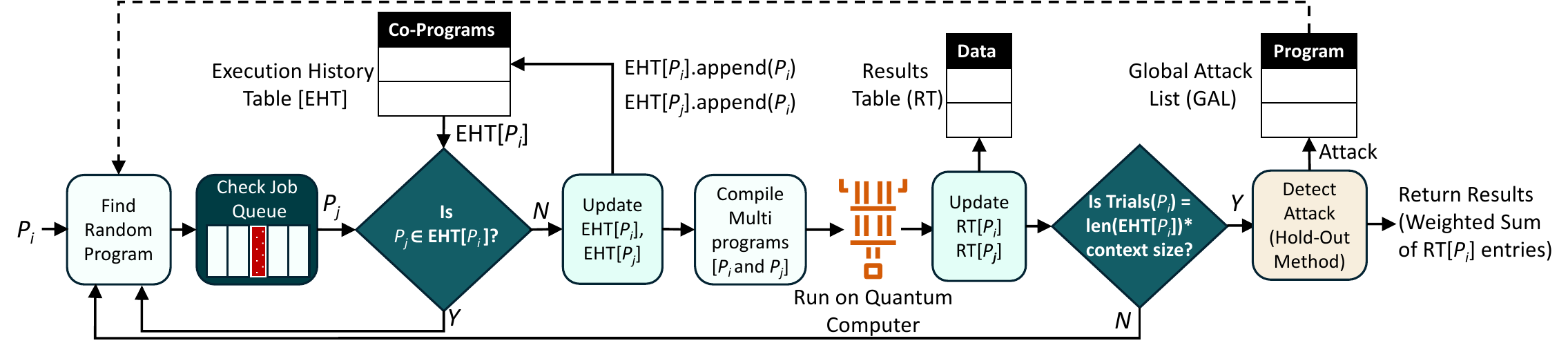} \vspace{-0.25in}
    \caption{Overview of \design{} with Attack Detection or \designenhanced{}.} \vspace{-0.15in}
\label{fig:design}
\end{figure*}

\section{\design{}: Design}
In this section, we describe the insights and implementation of our proposed design, \textit{\design{}: \designfullname{}}, that defends against \attackacro{}s. 
\subsection{Insight: Frequent Context Switching}

Existing multi-programming always co-locates the same programs throughout the execution. We refer to the simultaneous execution of a set of programs as a \textit{context}. 
Thus, programs corresponding to a victim and an attacker are always executed in a single context in default multi-programming. We identify this as a key reason for \attackacro{}s to be successful because all trials of the victim program are \textit{always} executed concurrently with the \attackacro{} program.  Our insight to defend against \attackacro{}s is to execute a program over multiple contexts instead of one while ensuring that it is executed with a unique program in each context. For example, assume a program $P$ must be executed for 8K trials. Existing multi-programming policies run 8K trials in one context. Thus, if $P$ is co-located with an \attackacro{} program $A$, all the 8K trials are vulnerable. \design{} overcomes this drawback by spreading the execution over multiple contexts. Thus, if $P$ is to be executed over eight contexts, each executing 1k trials, only one-eighth of $P$'s trials will now be vulnerable to the \attackacro{}, whereas the remaining seven contexts will remain unaffected (assuming they are co-located with benign programs). 

\subsection{Design Overview}
Figure~\ref{fig:design} shows an overview of \design{}.
It comprises an Execution History Table ($EHT$) and a Results Table ($RT$) to track program executions. The $EHT$ maintains an entry per program and tracks all other programs it has executed with. Thus, if a program $\texttt{P}_i$ is executed with $\texttt{P}_j$ in the first context, $EHT$[$\texttt{P}_i$] stores \{$\texttt{P}_j$\} and $EHT$[$\texttt{P}_j$] stores \{$\texttt{P}_i$\}. Now, if the program  $\texttt{P}_i$ is executed with $\texttt{P}_k$ in the second context, $EHT$[$\texttt{P}_i$] is updated to \{$\texttt{P}_j, \texttt{P}_k$\}, $EHT$[$\texttt{P}_j$] remains unchanged, and $EHT$[$\texttt{P}_k$] stores \{$\texttt{P}_i$\}. The $RT$ contains an entry per program which stores the output distributions from the contexts. Once a context finishes, the $RT$ entries corresponding to all the programs executed in the context are updated. In the above scenario, after the first context, the entries  $RT$[$\texttt{P}_i$] and $RT$[$\texttt{P}_j$] (initially empty) are updated. After the second context, $RT$[$\texttt{P}_i$] is appended with the output distribution obtained for $\texttt{P}_i$. Simultaneously, $RT$[$\texttt{P}_k$] is updated. Each program is allocated an entry (initially empty) in the $EHT$ and $RT$ when it enters the incoming job queue. The entries are removed only when all the trials requested by the program are executed and the results are returned to the user. \design{} also maintains a \textit{Global Attack List (GAL)} which is updated whenever an attack program is detected. To identify attack programs, \design{} compares the distributions from each context against each other and filters outlier candidates. 

To schedule a program $\texttt{P}_i$ in a shared environment over multiple contexts, \design{} uses the \textit{Multi-Programming with Frequent Context Switching (\algo{})} algorithm. The number of contexts required depends on the total number of trials to be executed for a program and the length of a context. The default implementation of \design{} uses eight contexts. By default, we execute 8K trials per program by default, a program is now be executed over eight contexts of 1K trials each. The \algo{} algorithm performs the following steps.  

\begin{enumerate}[leftmargin=0cm,itemindent=.5cm,labelwidth=\itemindent,labelsep=0cm,align=left, itemsep=0 cm, listparindent=0.5cm]
    \item \textit{Step-1}: The \algo{} algorithm finds a unique program from the incoming job queue ($Q$) to co-run $\texttt{P}_i$ with. Let $\texttt{P}_j$ be a potential candidate program ($\texttt{P}_j$ must not be in the $GAL$). 
    \item \textit{Step-2}: The \algo{} algorithm queries the $EHT$[$\texttt{P}_i$] to see if the list of co-programs $\texttt{P}_i$ was previously executed with include $\texttt{P}_j$, i.e, if program $P$ was ever previously executed with $\texttt{P}_j$. If $\texttt{P}_j \in EHT$[$\texttt{P}_i$], then $\texttt{P}_i$ cannot be executed with $\texttt{P}_j$ anymore and the algorithm goes back to \textit{Step-1} and find another candidate program. Otherwise, it proceeds to \textit{Step-3}. 
    \item \textit{Step-3}: The \algo{} algorithm updates the $EHT$ entries corresponding to both $\texttt{P}_i$ and $\texttt{P}_j$. These updates ensure that executions for both $\texttt{P}_i$ and $\texttt{P}_i$ are tracked and in future, when $\texttt{P}_j$ is scheduled, the trials executed with $\texttt{P}_i$ are accounted for. 
    \item \textit{Step-4}: The \algo{} algorithm compiles $\texttt{P}_i$ and $\texttt{P}_j$ for concurrent execution and runs them on the quantum computer. 
    \item \textit{Step-5}: The \algo{} algorithm updates $RT$[$\texttt{P}_i$] and $RT$[$\texttt{P}_j$]. 
    \item \textit{Step-6}:  The \algo{} algorithm checks if all the trials requested for  program $\texttt{P}_i$ has been completed or not. If completed, the results are analyzed to detect if any context executed an attack program. The entries $EHT$[$\texttt{P}_i$] and $RT$[$\texttt{P}_i$] are removed. Otherwise, the execution is repeated from \textit{Step-1}. 
    \item \textit{Step-7}: The distributions are analyzed using the \textit{Hold-Out method} to check if a context executed an attack program. 
    This is described in Section~\ref{sec:holdout} and referred to as \design{} with Attack Detection or \designenhanced{}.  
    If an attack is detected, the $GAL$ is updated. The final distribution of program $\texttt{P}_i$ is computed as a weighted sum of distributions from all contexts. The weights correspond to the estimated noise in each context and higher noise corresponds to lower weight. 
\end{enumerate}

\subsection{Attack Detection via Hold-Out Method}
\label{sec:holdout}
Despite context switching, a program may execute with \attackacro{}s, albeit with much lower probability. In this subsection, we discuss an attack detection scheme that further improves the performance of \design{}. 
Recollect that quantum programs produce both \textit{correct} and \textit{incorrect} outcomes/samples on real systems. The quality of the output distribution (say $D$) depends on the ratio of the correct to the incorrect outcomes, which we refer to as the \textit{CI Ratio} (higher is better). When a program executes with a \attackacro{}, increased noise levels lead to more incorrect outcomes than correct ones, reducing the CI Ratio. Thus, this distribution is significantly different from the one obtained by executing with a benign program, which has higher CI Ratio due to reduced noise levels. Although we cannot compute the CI Ratio because we do not know the correct outputs of programs, we can measure the divergence or distance between distributions from multiple contexts using statistical measures. Two distributions with similar CI Ratios will have much lower statistical distance than two distributions with dissimilar CI Ratios.

\begin{figure}[htp] 
\vspace{-0.1in}
    \centering
\includegraphics[width=\columnwidth]{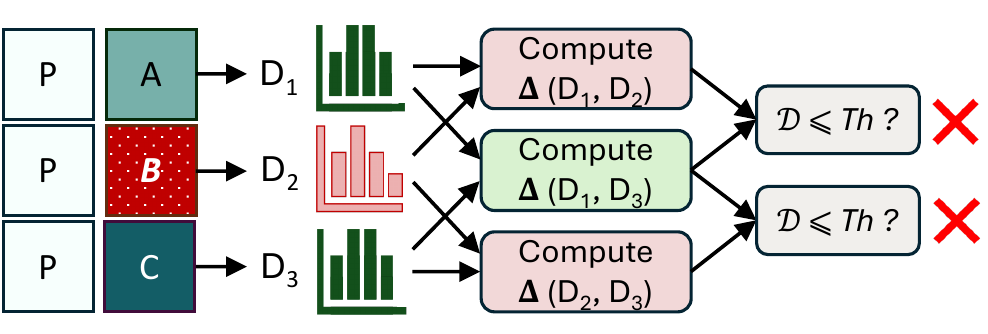} \vspace{-0.2in}
    \caption{Hold-Out method for detecting \attackacro{}s.} \vspace{-0.1in}
\label{fig:attackdetect}
\end{figure}

For example, if $P$ runs over three contexts with programs $A$, $B$, and $C$, we obtain distributions, $D_1$, $D_2$, and $D_3$, as shown in Figure~\ref{fig:attackdetect}.  Let $\Delta(i,j)$ be the Hellinger distance~\cite{hellingerdistance} between two distributions. $\Delta$ measures the statistical distance between two distributions and is bounded between 0 and 1, where 0 and 1 denote completely similar and dissimilar distributions. If $B$ belongs to an attacker, then $\Delta(D_1,D_2)$ and $\Delta(D_2,D_3)$ will be much higher than $\Delta(D_1,D_3)$ because both $D_1$ and $D_3$ are produced from contexts with benign programs and have comparable CI Ratios. In contrast, the CI Ratio of $D_2$ will be much lower than that of $D_1$ and $D_3$ due to increased levels of noise induced by $B$.  
Our studies show that Hellinger distance between samples of Di does not vary by more than $\approx0.3$ in the absence of attack programs. These slight variations result due to differences in error profiles of regions used in each context and differences in instruction patterns of the combined programs. We consider two distributions to be dissimilar if their $\Delta$ is 0.5 or above, whereas we consider them to be similar if $\Delta$ is below 0.35. \design{} identifies attacks by measuring the difference between $\Delta$ values, denoted by $\mathcal{D}$. If $\mathcal{D}$ exceeds an pre-defined \textit{Attack Detection Threshold ($Th$)}, \design{} classifies the context belonging to the common distribution between them as attacked. For example, in Figure~\ref{fig:attackdetect}, $B$ is identified as an attack because the $\Delta$s involving $D_2$ (top and bottom) far exceeds $\Delta(D_1,D_3)$.

To generalize for $C$ contexts, we propose the \textit{Hold-Out method}. This approach involves selecting a pair of contexts ([$i$, $j$]) and comparing against all other pairs involving $k \in {1,2,...C}, k \neq i \neq j$. If all $\Delta(i,k)$ and $\Delta(j,k)$ are low, $i$ and $j$ have comparable CI Ratio and are unlikely attack programs. If there are more $\Delta(i,k)$s that are low compared to many high $\Delta(j,k)$s, $j$ is detected as an attack using a majority vote, and the trials are discarded. We further explain this using an example from real IBMQ system in Section~\ref{sec:results}. The process is repeated for all context pairs.
To compute the final output distribution ($D$), the distributions from the non-attack contexts are merged by computing a weighted sum, where the weight of the $i^{th}$ context, $W_i$ is computed as $W_i$ = $\sum_{j=0}^{N}\Delta(D_i, D_j)$. $D$ is obtained from $\sum_{i=0}^{C}$ $\overline{W}_i*D_i$, where $\overline{W}_i$ denotes the normalized weight, given by $\overline{W}_i$ = $\frac{W_i}{\sum_{i=0}^{C} W_i}$. The $GAL$ is updated if any attack program is detected.

\subsection{Analysis of Resilience}
\label{sec:securityanalysis}
To measure the resilience (security), we use an analytical model. 
We assume $N$ programs out of which $K$ belongs to an attacker who is successful $\alpha$\% of the times. So, the probability that a program is attacked in multi-programming (as well as prior works QuCloud+, QuMC) is given by Equation~\eqref{eq:pattackbaseline}.

\vspace{-0.1in}
\begin{equation}
\textrm{P}_\textrm{baseline} = \frac{\alpha K}{N}
\label{eq:pattackbaseline}
\end{equation}

\subsubsection{Resilience levels}

We assume \design{} executes a program over $C$ contexts out of which $\beta$\% are run with programs of an attacker.
We assume two models of attack-

\begin{enumerate}[leftmargin=0cm,itemindent=.5cm,labelwidth=\itemindent,labelsep=0cm,align=left, itemsep=0.1 cm, listparindent=0.5cm]
    \item \textit{Strong attack}: At least 75\% of contexts run with attack programs.  Hence, there is a very high likelihood that the output of the program will be incorrect. In this scenario the program is the \textit{\textbf{least resilient}}. 
    \item \textit{Moderate attack}: At least 50\%  contexts are run with \attackacro{}s. So, there is a moderate likelihood that the program output may be incorrect and it is \textit{\textbf{moderately resilient}}.
\end{enumerate}

The probabilities that a program is attacked are denoted by $P^\textrm{strong}$ and $P^\textrm{moderate}$ for $\beta =75\%$ and 50\%, respectively. 

The probability that an attack is successful in a context is given by $\frac{\alpha K}{N}$, whereas the probability that a context runs with a benign application is ($1- \frac{K}{N}$). When attack programs are run in $\beta C$ contexts out of $C$, these contexts can be chosen in $\binom{C}{\beta C}$ ways. For example, if 50\% of 8 contexts run attack programs, the contexts can be chosen in $\binom{8}{4}$ ways. So, the probability that exactly $\beta C$ contexts are successfully attacked and the others run benign programs is given by Equation~\eqref{eq:exactattack}. 

\vspace{-0.1in}
\begin{equation}
\resizebox{0.90\hsize}{!}{$
\textrm{P}_{\textrm{exactly } \beta C\% \textrm{ attacked}} = \binom{C}{\beta C} \times (\frac{\alpha K}{N})^{\beta C} \times (1- \frac{K}{N})^{(C-\beta C)}$}
\label{eq:exactattack}
\end{equation}
Therefore, the probability that at least $\beta$\% of the contexts have been successfully attacked while the remaining contexts execute benign applications is given by Equation~\eqref{eq:atleastattack}. 

\vspace{-0.1in}
\begin{equation}
\resizebox{0.90\hsize}{!}{$
\textrm{P}_{\textrm{at least } \beta C\% \textrm{ attacked}} = \sum_{i=\beta C}^{C}  \binom{C}{\beta C} \times (\frac{\alpha K}{N})^{i} \times (1- \frac{K}{N})^{(C-i)}$}
\label{eq:atleastattack}
\end{equation}

\subsubsection{Results on Resilience} The probabilities of strong and moderate attacks, $P^\textrm{strong}$ and $P^\textrm{moderate}$ respectively, are computed using $\beta$ values of 75\% and 50\% in Equation~\eqref{eq:atleastattack}. Note that although \design{} runs each context with a unique program, the equations above provide a very good approximation when $N \gg C$, which is true in this case. 

\begin{figure}[htp]
    \centering \vspace{-0.1in}
\includegraphics[width=\columnwidth]{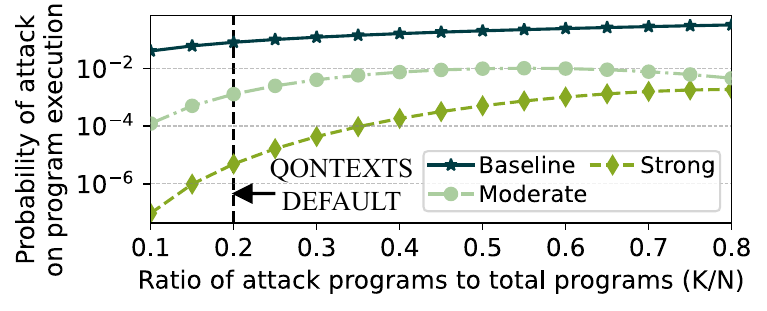}\vspace{-0.15in}
    \caption{Impact of increasing attack programs in queue.}
\label{fig:securityanalysisa} \vspace{-0.2in}
\end{figure}

Figure~\ref{fig:securityanalysisa} shows the probability that a program is attacked for increasing $\frac{K}{N}$, where $\frac{K}{N}$ is the ratio of the number of attack programs to the total number of programs. For our default analysis, we assume $N$=100 programs out of which $K$=20 belongs to the attacker, an attack is successful $\alpha$=40\% of the times (based on Section~\ref{sec:attackdemo}), and \design{} uses $C$=8 contexts. This results in $P_\textrm{baseline}$ as 8\%. For \design{}, $P^\textrm{moderate}$ = 0.13\% and $P^\textrm{strong}$ = $4.83\times 10^{-6}$. Thus, the resilience against a moderate and strong attack is $63\times$ and $16551\times$ compared to the baseline (more than three orders of magnitude). 

\subsubsection{Increasing contexts for higher resilience}
\label{sec:increase-contexts}
\begin{figure}[htp]
    \centering \vspace{-0.1in}
\includegraphics[width=\columnwidth]{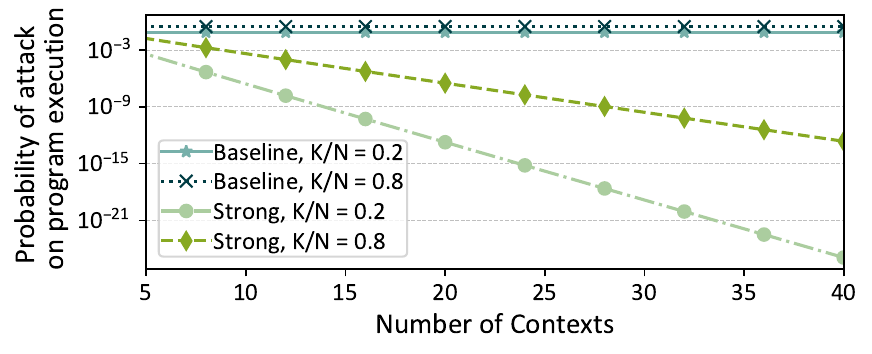} \vspace{-0.3in}
    \caption{Increasing contexts improves program resilience.} 
\label{fig:securityanalysisb}
\vspace{-0.2in}
\end{figure}

Figure~\ref{fig:securityanalysisb} shows the impact of increasing contexts on the probability that a program is strongly attacked for two $\frac{K}{N}$ ratios. In the baseline, a higher value of $\frac{K}{N}$ increases this probability, which remains constant, as expected. In \design{}, while the probability of attack also increases with increasing $\frac{K}{N}$, \design{} increases the resilience by increasing the number of contexts. For example, the probability that a program is strongly attacked when $\frac{K}{N}=0.2$ and $C=8$ is $4.83\times 10^{-6}$. \design{} achieves the same resilience for  increased $\frac{K}{N}=0.8$ by increasing the number of contexts to $C=17$ and making it tougher for an attacker to co-locate their program with a victim's for most of the execution.

\section{Evaluation Methodology}
We discuss the methodology used to evaluate our proposal. 

\subsection{Execution Framework}
By default, multi-programming runs two programs, as in prior works~\cite{das2019case, ash2020analysis, deshpande2023design, saki2021qubit, harper2024crosstalk}. We assume 20 benchmarks in a queue, 20\% of which belong to an attacker. We consider the following scheduling modes. 

\begin{enumerate}[leftmargin=0cm,itemindent=.5cm,labelwidth=\itemindent,labelsep=0cm,align=left, itemsep=0.1 cm, listparindent=0.5cm]
    \item \textit{\textbf{Isolated (Baseline)}}: Each program is executed in isolation. The mode denotes the highest achievable fidelity. 
    
    \item \textit{\textbf{Enhanced Multi-programming (}\textbf{EMP)}}: Two programs execute concurrently by default~\cite{das2019case} that are isolated via a layer of unused qubits~\cite{niu2023enabling} and no parallel CNOTs are run on links that are one hop from each other~\cite{liu2024qucloud+}. Thus, this mode is more secure compared to QuCloud+~\cite{liu2024qucloud+}. It represents the highest throughput achievable ($2\times$) and the fidelity of each program should be as close as possible to isolated execution. 
    
    \item \textbf{\textit{QuMC}}~\cite{niu2023enabling}: Two programs are run concurrently on qubit patches selected by the Qubit fidelity degree-based Heuristic Sub-graph Partition (QHSP) algorithm. A layer of unused qubits separate these regions to maintain one-hop isolation.

    \item \textbf{\textit{\design{}}}: Programs execute over $C$ contexts. This mode aims to achieve throughput comparable to EMP (without context switching) and fidelity comparable to isolated execution. We also use \textbf{\textit{\designenhanced{}}}. 
    By default, we use $C=8$ contexts because it corresponds to a very low probability of all contexts being strongly attacked (5 in a million). Our experiments confirm that this performs very well. Nonetheless, $C$ is a hyper-parameter that can be altered by the quantum service provider. For example, the provider can increase $C$ further to offer greater levels of resilience, as described in Section~\ref{sec:increase-contexts}. Alternately, the provider can estimate the percentage of attack programs by observing the insertion rate of programs into the global attack list and adjust the number of contexts in real-time.
\end{enumerate}

\subsection{Hardware: State-of-The-Art IBMQ Systems}
We use three IBMQ machines: 27-qubit IBM Hanoi, 127-qubit IBM Osaka, and 127-qubit IBM Sherbrooke. They employ tunable coupling and sparse heavy-hexagonal topologies for maximally reducing crosstalk via device-level improvements, enabling evaluations on already robust systems. 

\subsection{Benchmarks}
We choose benchmarks, shown in Table~\ref{tab:benchmarks}, from QASMBench~\cite{li2022qasmbench} and SupermarQ~\cite{tomesh2022supermarq} suites, consistent with prior works~\cite{smith2022scaling, nation2023suppressing, das2021lilliput, das2021adapt, patel2021robust, patel2022geyser, wang2023efficient, murali2019noise, murali2020software, murali2019full, das2023imitation}. We use programs corresponding to the attacker based on the method described in Section~\ref{sec:attackdemo}. 

\begin{table}[htb]
\centering
\vspace{-0.05in}
\begin{center}
\caption{Details of Benchmarks}
\vspace{-0.15in}
\label{tab:benchmarks}
\renewcommand{\arraystretch}{1.0}
\resizebox{\columnwidth}{!}{
\begin{tabular}{|c|c|c|c|}
\hline
Benchmark & Algorithm & \#Qubits& CNOTs \\
\hline
\hline
Adder & Adder~\cite{Cross_2022}& 10 & 65 \\ 
\hline
BV & Bernstein-Vazirani~\cite{bernstein1993quantum} & 11 & 6 \\
\hline
Dnn & Neural Network~\cite{stein2021hybrid} & 8 & 192 \\ 
\hline
GHZ & Bell-state~\cite{GHZ} & 9 & 8 \\ 
\hline
HS & Hamiltonian Sim~\cite{campbell2019random}& 10 & 18 \\ 
\hline
Ising & Ising Model~\cite{ising1925beitrag}& 10 & 90 \\ 
\hline
QAOA & Maxcut with p=1~\cite{farhi2014quantum}& 10 & 135 \\ 
\hline
QPE & Phase Estimation~\cite{kitaev1995quantum} & 9 & 43 \\ 
\hline
SAT & Optimization~\cite{Cross_2022}& 11 & 252 \\   
\hline
\end{tabular}}
\vspace{-0.2in}
\end{center}
\end{table}

\subsection{Figure-of-Merit}
\label{sec:figure-of-merit}
\noindent \textbf{Attack Success Criterion:} We consider an attack successful if (1)~the correct answer can be determined during isolated execution but cannot be determined while multiprogramming (for programs with one correct output) or (2)~if the fidelity while multiprogramming is reduced by more than 12\% compared to isolated mode (for programs with distributions as output). We accept up to 12\% lower fidelity while multiprogramming because the latter is known to reduce fidelity and this threshold is based on prior works~\cite{das2019case,deshpande2023design,niu2023enabling}.

\vspace{0.05in}
\noindent \textbf{Throughput:} We measure \textbf{\textit{throughput}} using Equation~\eqref{eq:throughput} as the ratio of the total latency of a program in isolated mode to the latency in a given multi-programming mode. 

\vspace{-0.1in}
\begin{equation}
\textrm{Throughput} = \frac{\textrm{Latency of a program in isolated mode}}{\textrm{Latency of a programs in a given mode}}
\label{eq:throughput}
\end{equation}

\vspace{0.05in}
\noindent \textbf{Fidelity:} We measure \textit{\textbf{fidelity}} using Total Variation Distance~\cite{TVD} between the noise-free output distribution on a simulator ($P$) and the noisy distribution from a real device ($Q$), as shown in Equation~\eqref{eq:fidelity}. This metric is derived from various prior works~\cite{patel2021robust,jigsaw,adapt,ash2020experimental,sarovar2020detecting}.
Fidelity ranges between $0$ and $1$, where $1$ represents completely identical distributions. 

\vspace{-0.1in}
\begin{equation}
\textrm{Fidelity} = 1- \sum_{i=1}^{k}\mid \mid P_i - Q_i \mid \mid 
\label{eq:fidelity}
\end{equation}

\noindent\textit{Ideally, higher fidelity, throughput, and security are desirable. } 

\begin{figure*}[htp]
    \centering
\includegraphics[width=0.95\textwidth]{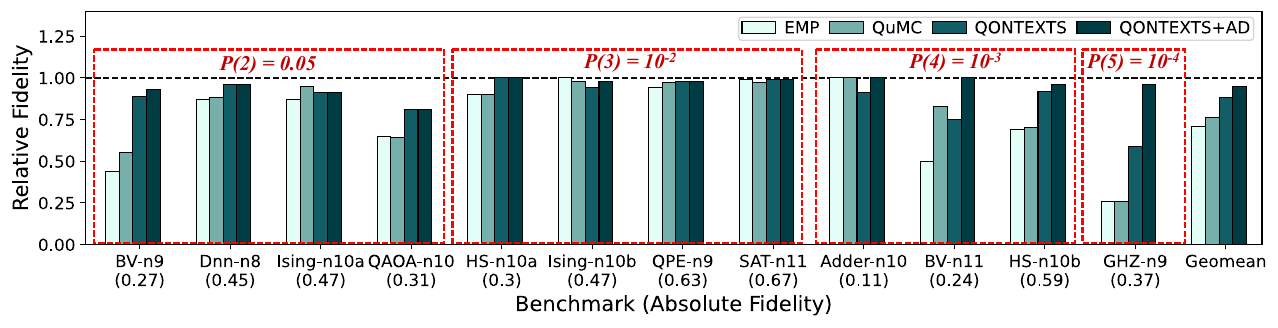} \vspace{-0.15in}
    \caption{Fidelity of  multi-programming modes relative to isolated execution. Data collected from experiments on real hardware: IBM-Hanoi, IBM-Osaka, and IBM-Sherbrooke. Here, $P(N)$ denotes the probability that $N$ contexts are attacked. Due to space constraints, we only shows cases where there are at least two of more (up to five) contexts are attacked in \design{}.} \vspace{-0.15in}
\label{fig:results}
\end{figure*}

\section{Results}
\label{sec:results}
In this section, we discuss the performance of \design{}. 

\subsection{Security}
Table~\ref{tab:security} shows the security of various modes. \designenhanced{} is the most secure and offers up to three orders of magnitude higher resilience than EMP and QuMC (Section~\ref{sec:securityanalysis}). %

\begin{table}[htb]
\centering
\begin{center}
\caption{Security against \attackacro{}s of different policies}
\vspace{-0.1in}
\label{tab:security}
\renewcommand{\arraystretch}{1.1}
\setlength{\tabcolsep}{3pt}
\begin{tabular}{|c|c|c|c|c|}
\hline
Benchmark & EMP & QuMC & \design{} & \specialcell{\design{}\\+AD} \\
\hline
BV-n9  & \redcross & \redcross & \greentick & \greentick\\
\hline
Dnn-n8  & \redcross & \greentick & \greentick & \greentick \\
\hline
Ising-n10a  &\redcross & \greentick & \greentick & \greentick \\
\hline
QAOA-n10  & \redcross & \redcross &\greentick & \greentick  \\
\hline
HS-n10a  & \redcross & \redcross & \greentick & \greentick \\
\hline
Ising-n10b  & \greentick & \greentick & \greentick & \greentick \\
\hline
QPE-n9  & \redcross & \greentick &\greentick & \greentick \\
\hline
SAT-n11 & \greentick & \greentick &\greentick & \greentick \\
\hline
Adder-n10  &\greentick & \greentick & \redcross & \greentick \\
\hline
BV-n11  & \redcross & \greentick  & \redcross & \greentick \\
\hline
HS-n10b  &\redcross & \redcross & \greentick & \greentick \\
\hline
GHZ-n9  & \redcross & \redcross & \redcross & \greentick \\
\hline
\end{tabular}
\end{center}
\begin{tablenotes}
        \item[a]\textit{$^*$Note:  Due to space constraints, we only show results for cases where at least one context includes an attack program.}
    \end{tablenotes}
\vspace{-0.15in}
\end{table}

\noindent \textbf{\textit{Demonstration of strong attack:}} The Adder-n10 program shows a case in which it runs with benign programs in EMP and is secure. QuMC runs it with \attackacro{} but is able to defend successfully. Here, \design{} is specifically crafted to demonstrate a strong attack scenario and 50\% of the contexts are forced to include \attackacro{}s (no random selection). \designenhanced{} successfully defends even in such cases, highlighting it superiority in enabling secure multi-programming. As already discussed in Section~\ref{sec:securityanalysis}, in practice, the probability of encountering this scenario is very low.

\subsection{Throughput}
\begin{figure}[htp]
    \centering
    \vspace{-0.1in}
\includegraphics[width=\columnwidth]{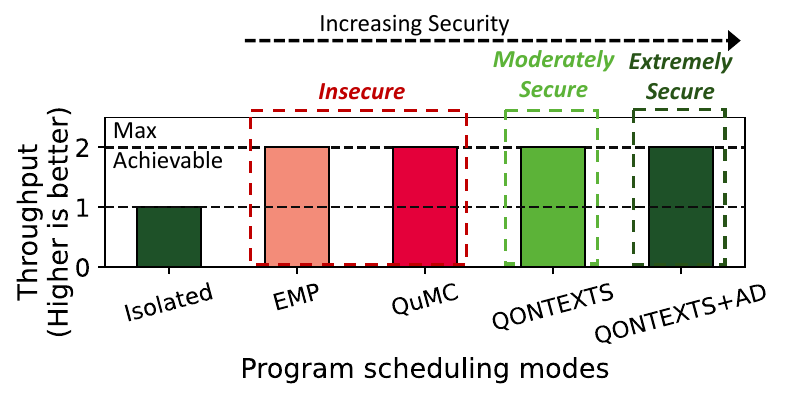}  
    \vspace{-0.25in}
    \caption{Throughput of scheduling modes. \design{} is secure and maximizes throughput, unlike prior works.}%
\label{fig:throughput}  \vspace{-0.15in}
\end{figure}

Figure~\ref{fig:throughput} shows throughput of various scheduling approaches. The isolated execution mode offers the highest fidelity but no throuhgput improvements. %
Both EMP and QuMC achieves the  maximum attainable throughput ($2\times$ in our default setting) but are not secure. \design{} achieves the same throughput as EMP and QuMC but is relatively more secure because only a fraction of the trials are now executed with attack programs. However, it may still reduce the fidelity of programs when too many contexts include attack programs. In contrast,  
\designenhanced{} achieves the same throughput while remaining as secure as isolated execution because it detects the attacked contexts and excludes them from computing the final output distributions.  

\subsection{Fidelity}
Figure~\ref{fig:results} shows the fidelity of benchmarks relative to isolated mode when shared environments include attack programs.     
\design{} and \designenhanced{} improves fidelity by $1.28\times$ and $1.33\times$  on average compared to EMP respectively and achieves identical fidelity as isolation in the best case. \design{} has higher fidelity than EMP because victim programs are run with the attack programs for all the trials in EMP, whereas in \design{}, programs run over multiple contexts along with benign circuits, thus reducing the number of trials impacted by attacks or errors. %

\subsection{Example of Attack Detection in \designenhanced{}}

\begin{figure}[htp]
    \centering
\includegraphics[width=\columnwidth]{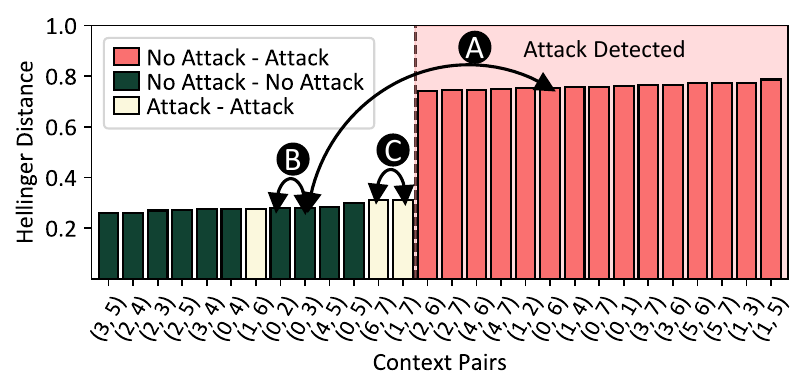} %
    \caption{Difference in Hellinger distances between distributions from different contexts enables us to detect attacks.} \vspace{-0.15in}
\label{fig:ghzattackdetection}
\end{figure}

The \textit{Hold-Out method} detects attacks by comparing the distance between the distributions from $B$ contexts when paired with each other. Figure~\ref{fig:ghzattackdetection} shows the distance between the distributions of the GHZ\_n10 benchmark, in which three out of the eight contexts (1, 6, and 7) runs attack programs. The eight distributions of the program can be paired in $^8C_2$ or $\binom{8}{2}$ ways, yielding 24 pairs. We notice up to $\sim$77\% higher distance when a context includes an attack compared to a benign one. Let $\Delta(i,j)$ be the distance between the distributions from contexts $i$ and $j$. We observe that-  \bcircled{A}: $\Delta(0,3)$ is way lower than $\Delta(0,6)$; whereas \bcircled{B}: $\Delta(0,2)$ and $\Delta(0,3)$ are low and comparable. Note that the distance between distributions from two contexts that both run attack programs would be low because they are both equally inaccurate and have similar noisiness. For example, \bcircled{C}: $\Delta(6,7)$ and $\Delta(1,7)$ have low distances. Attacks can be identified by observing the distances. For example, to identify whether 0 or 6 is an attack or not, we look at the $\Delta$ between 0 and all other contexts, and $\Delta$ between 6 and all other contexts. We observe two scenarios ($\Delta(1,6)$, $\Delta(6,7)$) where context 6 has a low distance while many more scenarios exist where context 0 has a low distance ($\Delta(0,2)$, $\Delta(0,3)$, $\Delta(0,4)$ and $\Delta(0,5)$). 

\subsection{Impact of Context Switching Overheads}
\label{sec:latencyimpact}
\subsubsection{Impact on Individual Program Latency}
The end-to-end latency of a program is the sum of (1)~\textit{queuing time} which is the time spent in the queue waiting to access a machine, (2)~\textit{program execution time} which is the sum of the time taken to load the program on to the control FPGAs, and the time spent in running the circuit. 
Typically, queuing times range between few hours to several days~\cite{ma2024understanding}. Loading latencies, denoted by $t_{load}$, is proprietary data confidential to device providers. We run benchmarking circuits to reverse engineer this timing from Qiskit Runtime. Table~\ref{tab:loadingoverhead} shows the runtime for different number of programs and trials, while the total number of trials executed altogether remains constant. We obtain a similar latency, within a 0.4-second range, for all settings, indicating that the loading latency is constant regardless of the number of programs loaded on IBM devices. The time taken to run the circuit on the quantum hardware is often a few milliseconds (assuming superconducting systems and a few thousands of trials). Thus, queuing times far exceed the total program execution time. Moreover, the program execution time largely remains unaffected in \design{} because context switching incurs negligible overheads.

    \begin{table}[!htb]
    \vspace{-0.1in}
        \centering
        \begin{center}
        \caption{Runtime variation with program counts and trials}
        \vspace{-0.15in}
        \label{tab:loadingoverhead}
        \renewcommand{\arraystretch}{1.0}
        \begin{tabular}{|c|c|c|}
            \hline
            Program Count & Trials & Latency (s)\\
            \hline
            \hline
            1 & 8K & 5.15 \\
            \hline
            2 & 4K & 5.50 \\
            \hline
            4 & 2K & 5.46\\
            \hline
            8 & 1K & 5.53 \\
            \hline
        \end{tabular}
        \vspace{-0.1in}
        \end{center}
    \end{table}

\subsubsection{Impact on Throughput}
To compute the impact on throughput, we compute program execution time using an analytical model. We assume a program runs \texttt{T} trials, latency per trial is $t_{trial}$, a repetition delay time of $t_{wait}$ between trials, and $t_{load}$ is program loading time onto the control FPGAs. During context switching, $t_{load}$ and $t_{wait}$ overlap. So, the context switching latency, $t_{switch}$, is the maximum of the two. Let $S$ programs run concurrently during multi-programming. The latency of a program in isolated and multi-programming modes are given by Equations~\eqref{eq:baselinelatency} and~\eqref{eq:multiprogrammingtime} respectively.

\begin{equation}
\tau_\textrm{isolated} =  t_{load} + \texttt{T} \times t_{trial} + (\texttt{T}-1) \times t_{wait}
\label{eq:baselinelatency}
\end{equation}

\begin{equation}
\resizebox{0.90\hsize}{!}{$
\tau_\textrm{multi-programming} =  \frac{t_{load} + \texttt{T} \times t_{trial} + (\texttt{T}-1) \times t_{wait}}{S}$}
\label{eq:multiprogrammingtime}
\end{equation}

Assuming $C$ contexts are used in \design{}, the execution time of a program is given by Equation~\eqref{eq:designtime}. 

\begin{equation}
    \label{eq:designtime}
    \resizebox{0.90\hsize}{!}{$
    \tau_\textrm{\design{}}= \frac{C \times t_{switch} + \texttt{T} \times t_{trial} + (\texttt{T}-1) \times t_{wait}}{S}$}
\end{equation}

\begin{figure}[htp]
    \centering
\includegraphics[width=\columnwidth]{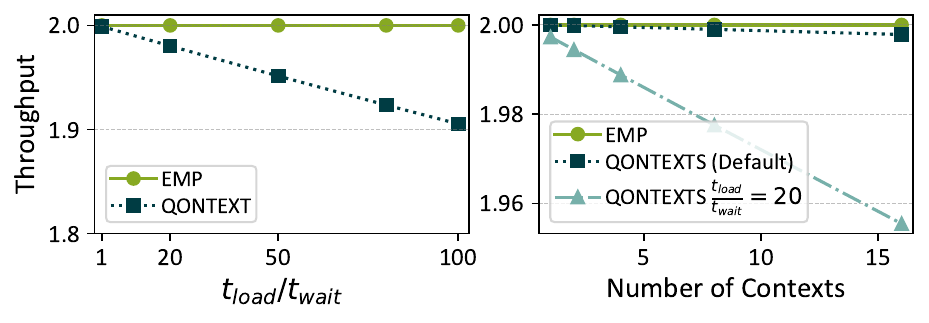} \vspace{-0.2in}
    \caption{Throughput for (a) $\frac{t_{load}}{t_{wait}}$ and (b) contexts.} 
\label{fig:speedupfigure}
\end{figure}

For generalization, we assume $t_{wait}$=250 $\mu$s, the default on IBM systems, \texttt{T}=10K trials, $t_{trial}$=100 $\mu$s, and $S$=2. Figure~\ref{fig:speedupfigure} shows throughput for- (1)~different $\frac{t_{load}}{t_{wait}}$ with default contexts $C$=8, and (2)~variable contexts with fixed $\frac{t_{load}}{t_{wait}}$=1 and 20. \design{} has identical throughput as EMP. Moreover, increasing the number of contexts does not degrade throughput significantly because loading latencies are negligible.

\subsubsection{Impact on Quality of Services}

To evaluate the impact of quality of services at the service provider level, we conduct a queue simulation and analyze the per-job completion time under various system loads. The system load is characterized by the job arrival rate, defined as the number of new jobs arriving during one job's execution time. We compare \design{} with isolated execution and EMP. Given that both EMP and \design{} require multiple circuit executions, we simulate scenarios with system loads ranging from two job arrivals per execution time to 10 jobs, though in practice the load is often substantially higher~\cite{ravi2021quantum}. The results show that the performance gap between EMP and \design{} narrows significantly as system load increases, with both methods maintaining considerably shorter completion times compared to isolated execution. The gap between EMP and \design{} is almost negligible compared to that between isolated and \design{}.

\begin{figure}[htp]
    \centering
    \includegraphics[width=\linewidth]{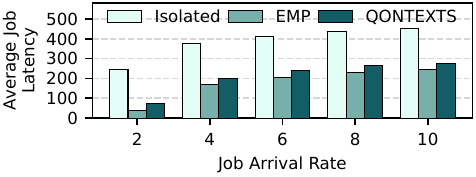}\vspace{-0.1in}
    \caption{Average job latency vs. system load. \design{} demonstrates similar scaling as compared to EMP, while both significantly outperform isolated execution.}
    \vspace{-0.10in}
    \label{fig:queue_sim_results}
\end{figure}

\subsection{Scalability Analysis}
\label{sec:scalability}
We discuss scalability across three vectors- (1)~applicability, (2)~concurrency, and (3)~program sizes. 

\subsubsection{\textbf{Applicability}} As context switching overheads are negligible and the design structures (such as tracking tables) incur nominal overheads, \design{} is scalable in terms of applicability because it can be seamlessly integrated in existing software stacks, enabling practical adoption.

\subsubsection{\textbf{Concurrency}} To study the scalability in terms of number of concurrent programs and \attackacro{}s, we conduct additional studies using IBMQ Sherbrooke. We co-locate a \attackacro{} with three other programs (QAOA, HS10 and GHZ9). In the default multi-programming (\textit{EMP}), ZKTAs successfully reduce fidelity by 29.7\%, whereas \textbf{\textit{\design{} achieves $4\times$ throughput and remains secure.}}

We further increase the concurrency to 7. ZKTAs continue to succeed, reducing fidelity by 21.2\% In contrast, \textbf{\textit{\design{} offers $7\times$ throughput while remaining secure.}}

\subsubsection{\textbf{Program Sizes}} We study \textbf{programs with up to 20 qubits} (BV19, QFT18, and DNN16). Programs beyond this size yield extremely noisy distributions even in isolation (fidelity < 0.02\%) and cannot be meaningfully used. \attackacro{}s successfully lower fidelity by 21.4\% on average and \textit{\textbf{\design{} strongly defends}} against them. 

We also run a \attackacro{} with programs of non-uniform sizes (Adder-n10, Pea-n5, HHL-n7). The \attackacro{} successfully reduce the fidelities of Pea-n5 by 63.6\%. In contrast, \textit{\textbf{\design{} remains secure even for program of uneven sizes}}.

\section{Discussion}
\subsection{Why is Discarding Noisy Contexts Fine?}
\label{sec:detection_scheme}

Contexts with \attackacro{}s yield fewer correct outcomes and huge number of incorrect outcomes due to higher noise levels, reducing the CI (correct to incorrect outcomes) Ratios. To handle attacked contexts, we have three options: (1)~include their noisy results in the final distribution, (2)~discard the compromised runs, or (3)~re-execute the program for more trials to compensate for the discarded runs. The first options yields output distributions with moderate CI Ratios because it combines results with both high and low CI Ratios (thereby averaging out because the probabilities of incorrect outcomes increase, whereas that of the correct outcomes are attenuated). In contrast, the second approach only combines results from contexts with high CI Ratios, yielding a much more accurate distribution because the high CI Ratios accentuate the correct outcomes. 

\begin{figure}[htp]
    \centering
\includegraphics[width=1.0\columnwidth]{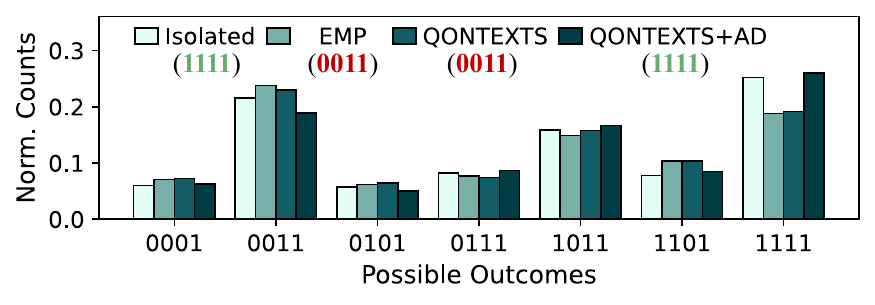}  \vspace{-0.25in} 
    \caption{Output distribution for the 5-qubit PEA benchmark, with P(outcomes) < 2\% not shown for clarity.}
    \vspace{-0.1in}
\label{fig:normdist}
\end{figure}

To show this, we run a 5-qubit PEA program~\cite{li2022qasmbench} where 4 out of 8 contexts run \attackacro{}s. CI Ratios are 0.33, 0.23, and 0.24 for isolated mode, EMP, and \design{} respectively. \design{} represents the first approach from above. The correct output appears with a probability of 25\% and can be identified in isolated mode. This probability reduces to 18.8\% and 19.2\% for EMP and \design{} respectively, incorrect producing \textit{\textbf{0011}} as the program output (see Figure~\ref{fig:normdist}). Contexts with \attackacro{}s have CI Ratios of 0.02 (\textit{too low}), 0.18, 0.18, and 0.19, compared to 0.34 on average for the benign ones. When these noisy contexts are used in the aggregated results, the overall CI Ratio is only 0.24. In contrast, \designenhanced{} discards the compromised runs, yielding CI Ratio of 0.35 and a 26\% probability for the correct output. Now, the correct output \textit{\textbf{1111}} can be inferred.

We adopt the second approach because our studies show that it performs similar to the third one (as newer benign contexts yield similar CI Ratios). Nonetheless, service providers may also choose the latter because it does not lower the throughput as only a few trials need to be re-executed and \designenhanced{} offers high resilience.

\subsection{Attack Detection Threshold Selection}
\label{sec:detection_scheme}
\begin{figure}[htp]
    \centering

\includegraphics[width=1.0\columnwidth]{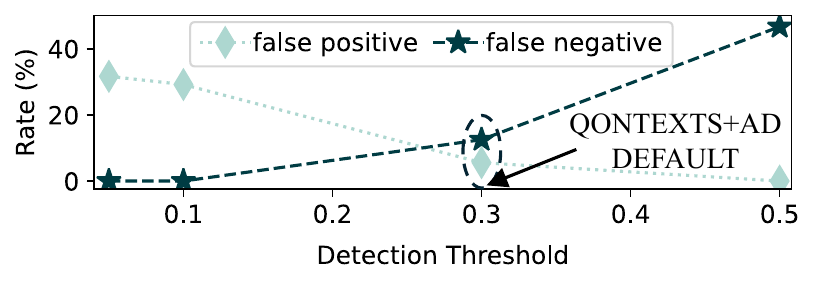} 
\vspace{-0.2in}
    \caption{False positive and False negative rate of the detection scheme with increasing classification Threshold} %
\label{fig:falseposneg}
\end{figure}

The selection of the \textit{Attack Detection Threshold ($Th$)} is crucial in \designenhanced{}. Low thresholds cause more contexts to get incorrectly classified as attacks because it flags even minor deviations as potential threats. This causes high \textbf{false positive} rates that decreases as the threshold increases. In contrast, high thresholds cause more attack contexts to be misclassified as benign, causing \textbf{false negative} rates that increase with thresholds. Based on our studies (Figure~\ref{fig:falseposneg}), we choose 0.3 as the default threshold, achieving a false positive rate of 5.56\% and a false negative rate of 12.5\%, aligning with Bradley's liberal criterion~\cite{ferrer2023comparison}. Service providers may tune this parameter as needed. 

\section{\design{} for Variational Algorithms}
\label{sec:zktaonvqas}
\begin{figure}[htp]
\vspace{-0.1in}
    \centering
\includegraphics[width=\columnwidth]{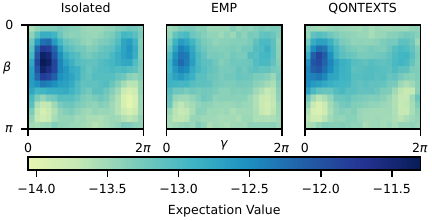} \vspace{-0.25in}
    \caption{Landscapes of a QAOA MaxCut program on real IBM Hanoi for isolated, EMP, and \design{} modes.} \vspace{-0.1in}
\label{fig:qaoalandscape}
\end{figure}

Near-term applications use variational quantum algorithms (VQAs)~\cite{mcclean2016theory,biamonte2017quantum,peruzzo2014variational,farhi2014quantum} that train a parametric circuit over many iterations. The expectation value of the distribution from each iteration is used to tune the circuit parameters for the next iteration, until the optimization converges and the optimal parameters are found. The distribution of the optimal circuit is used to find the program output. The performance of VQAs depends on the ability to perform gradient descent on the optimization landscape. %
We study the impact of \attackacro{}s on VQAs. Figure~\ref{fig:qaoalandscape} shows the landscapes of an 8-qubit QAOA for MaxCut problem on IBM Hanoi. The circuit has two parameters $\beta,\gamma$ and each point on the landscape shows the expectation value for a combination of $\gamma$ and $\beta$.  

Due to limited hardware access, we use three contexts for \design{}. In EMP, each QAOA program is run with an attack program, whereas in \design{}, it is run with a benign and two attack programs. We observe \attackacro{}s significantly degrade the sharpness of the landscapes, reducing gradients, whereas the landscape from \design{} is  sharper. 
\section{Related Work}
\design{} is orthogonal to most prior works on multi-programming~\cite{niu2023enabling,niu2022parallel,liu2021qucloud,ohkura2022simultaneous,niu2022multi,10.1145/3470496.3527434,mineh2023accelerating,resch2021accelerating,park2023quantum} and multi-system execution~\cite{ravi2021adaptive,yao2022mtmc,micro1,wang2024qoncord} that focuses on fair resource allocation and instruction scheduling.
Multi-programmed systems can be attacked by leveraging various sources of errors. \design{} focuses on crosstalk-based attacks~\cite{choudhury2024crosstalk}, similar to QuCloud+~\cite{liu2024qucloud+} and QuMC~\cite{niu2023enabling} but is more secure than them. Saki et al. investigate crosstalk based outcome corruption attacks~\cite{ash2020analysis} using one hop away CNOTs. However, our studies show that these attacks extend beyond just closely positioned CNOT links. Moreover, \design{} is secure against these attacks too. 

The qubit-sensing attack~\cite{saki2021qubit} exploits readout error bias in which state $\ket{1}$ is more vulnerable to errors than state $\ket{0}$ and the measurement outcome of a qubit depends on the outcome of other qubits. This attack exploits measurement bias to sense victim program outcomes using malicious circuits. It requires exhaustive profiling to assess the bias and craft attack circuits which limits scalability. \design{} can defend against such attacks too because it randomizes the co-running programs, making it harder for the attacker.

Quantum systems are vulnerable to attacks even if not multi-programmed~\cite{trochatosdynamic,trochatos2023quantum,trochatos2023hardware,ash2020analysis, abrams2019methods, upadhyay2022robust, lu2020quantum, saki2021qubit, xu2023exploration,saki2021split}. For example, fast and accurate fingerprinting reveals proprietary information about systems that are otherwise unknown~\cite{mi2021short,morris2023fingerprinting,smith2023fast}. 
Qubit resets can also be exploited for attacks. When qubits are reset at the end of a program execution, its outcome can be inferred by the next program~\cite{tan2023extending,xu2024thorough,xu2023securing}.  
Using random single-qubit gates or one-time-pads before resets can defend against these attacks~\cite{xu2024thorough,xu2023securing}. 
Program outputs can also be inferred by studying power traces of controllers used to generate control pulses~\cite{xu2023exploration,erata2024quantum}. \design{} is orthogonal to these works.

\section{Conclusion}
Crosstalk-induced errors can be leveraged to attack multi-programmed quantum systems. 
We propose \textit{\design{}, \designfullname{}}, that alleviates such attacks by splitting a program execution over multiple \textit{contexts}, in each of which it is run concurrently with a unique program for a subset of the trials. Thus, now only a fraction of the program execution is vulnerable to attacks, while the other contexts run successfully. We enhance \design{} further by comparing the distance between distributions from different contexts to detect attacks. \design{} with attack detection, \designenhanced{}, alleviates crosstalk-based attacks and increases program resilience by up to three orders of magnitude, while improving fidelity by 1.33x on average compared to multi-programming and achieves the highest attainable fidelity (equivalent to isolated mode) in the best-case. 

\section*{Acknowledgments}
We thank the reviewers of MICRO-2024, ASPLOS-2025, and ISCA-2025 for their comments and feedback. We thank Nicolas Delfosse for technical discussions on the security analysis. We thank the IBM Quantum Credits Program for offering us access to some recent IBMQ hardware. Poulami Das acknowledges the support through the AMD endowment at UT Austin. Prashant J. Nair and Meng Wang are supported by NRC Canada grants AQC 003 and AQC 213, and Natural Sciences and Engineering Research Council of Canada (NSERC) [funding number RGPIN-2019-05059] for this work. This material is based upon work supported by the U.S. Department of Energy, Office of Science, National Quantum Information Science Research Centers, Co-design Center for Quantum Advantage (C2QA) under contract number DE-SC0012704, (Basic Energy Sciences, PNNL FWP 76274). This research used resources of the Oak Ridge Leadership Computing Facility, which is a DOE Office of Science User Facility supported under Contract DE-AC05-00OR22725.

\section*{APPENDIX}
\begin{appendix}
Here, we discuss the \algo{} used in \design{}.
\begin{algorithm}
\caption{\algo{} used in \design{}}\label{alg:cap}
\textbf{Input:} Program ($\texttt{P}_i$), Execution History Table ($EHT$), Results Table ($RT$), Trials ($T$), Job Queue ($Q$), Global Attack List ($GAL$)\\
\textbf{Output: } Output Distribution of $\texttt{P}_i$ ($D$) 
\begin{algorithmic}[1]
\Function{MFCS}{$P, Q, EHT, RT$} %
\State $N \gets \frac{T}{\textrm{Context Size}}$
\phantom{xxx}\color{blue}// Run $P$ over $N$ contexts \color{black}

\While{$\text{len}(EHT[\texttt{P}_i]) \neq N$}
    \State $\text{Randomly select program } \texttt{P}_j \text{ from } Q$
    \If{$\texttt{P}_j \notin EHT[\texttt{P}_i] $ and $\texttt{P}_j \notin GAL$ }
        \State $EHT[\texttt{P}_i].\text{append}(\texttt{P}_j)$
        \State $EHT[\texttt{P}_j].\text{append}(\texttt{P}_i)$
        \State Compile $[\texttt{P}_i, \texttt{P}_j]$ and execute concurrently
        \State Update $RT[\texttt{P}_i]$, $RT[\texttt{P}_j]$
    \EndIf
\EndWhile

\State $D \gets \textrm{Weighted }{RT[\texttt{P}_i]} \textrm{ post attack detection}$  
\State Remove $EHT[\texttt{P}_i]$, $RT[\texttt{P}_i]$, Update $GAL$
\State \textbf{return} $D$ \color{blue}// Return output distribution to user \color{black}
\EndFunction
\end{algorithmic}
\end{algorithm}
\end{appendix}

\bibliographystyle{unsrt} %
\bibliography{references}

\end{document}